\documentclass[11pt]{article}
\usepackage[T1]{fontenc}
\usepackage[margin=1in]{geometry}
\usepackage{lmodern,microtype,amsmath,booktabs,graphicx,xcolor}
\usepackage[colorlinks=true,linkcolor=black,citecolor=teal!60!black,urlcolor=teal!60!black]{hyperref}
\usepackage{xurl}
\hypersetup{pdftitle={Vetoes and missed agreements in language-model bargaining},pdfauthor={Fuad Ali}}
\title{\bfseries Vetoes and missed agreements in language-model bargaining}
\author{Fuad Ali}
\date{November 2026}
\begin{document}
\maketitle
\begin{abstract}
A language model played three representatives, each instructed to get the best payoff for its own group. One representative requested a plan, another chose the sole offer, and all three voted. If anyone voted no, each group received its fallback payoff. In 60 games with Qwen3-Next-80B, the representative at risk of a loss always requested a plan that would leave everyone better off. In each game, two copies of the proposer saw the same payoffs, peer request and reminder that the offer needed three yes votes. Showing the proposer how many groups gained under each plan raised mutually beneficial agreements from 6/60 to 31/60 (paired increase 41.7 percentage points, conservative 95\% interval 18.8 to 58.5). Two of the four plans benefited everyone, so picking uniformly would put a mutual gain on the ballot half the time on average. In an earlier study, GPT-4.1 played 12 conflict games under two message conditions. Majority rule passed 15 of 24 offers, all at one group's expense. Unanimity passed two, both mutual gains. In the Qwen games, voters never saw the counts. They approved all 37 offers that helped everyone. On all 83 offers that would have cost one group points, its representative voted no.
\end{abstract}

\section{What got put to a vote}

In one game the three constituencies would receive 41, 42 and 42 points if the vote failed. The second representative asked for a plan paying 51, 58 and 60. The first representative, who controlled the ballot, offered 77, 33 and 71 instead. The second voted no, the others voted yes, and everyone received the fallback. The same thing happened when the proposer saw a count of how many constituencies each plan would benefit (Table~\ref{tab:menu}).

\begin{table}[htbp]\centering\small
\caption{Payoffs in case 00 of the matched-reminder experiment. Only the proposer in the count condition saw the last column.}
\label{tab:menu}
\begin{tabular}{lrrrr}\toprule
Plan & Proposer & Requesting party & Third party & Number gaining\\\midrule
P1 (requested) & 51&58&60&3\\
P2 (offered) & 77&33&71&2\\
P3 (fallback) & 41&42&42&0\\
P4 & 57&47&50&3\\\bottomrule
\end{tabular}
\end{table}

A record of the vote would tell us that the second party blocked the deal. The record of the request shows that the same party had proposed a deal all three had reason to approve. The proposer even had an alternative, P4, that paid its own constituency six points more than the requested plan and still left the other two ahead. Neither compromise reached the ballot.

In human three-person bargaining experiments, unanimity can delay agreement, and changing the procedure for making offers can change who benefits \cite{tremewan}. Language-model negotiation studies have measured how bargaining tactics affect payoffs \cite{bianchi} and how agents handle multi-party agreements with competing interests \cite{abdelnabi}. RoundTable compares voting procedures in multi-agent collaboration \cite{cho}. A final payoff tells us who gained, and a ballot tells us who blocked an offer. Neither tells us whether the voter had already put forward a compromise. We follow each request through the proposer's choice and three binding ballots. In the final experiment the party facing a loss asked for a mutual gain in all 60 games. Comparing each request with the subsequent offer shows which compromises never reached the voters.

\section{Games and comparisons}

Each action comes from a separate call to the same model within a study. Representatives are instructed to maximize their own constituency's payoff. The \emph{agenda setter} chooses the sole final offer. All three representatives then vote. For plan \(p\), let \(u_i(p)\) denote constituency \(i\)'s payoff and let \(s\) denote the fallback. The number shown to the proposer in the counts condition is
\begin{equation}
G(p)=\sum_{i=1}^{3}\mathbf{1}\{u_i(p)>u_i(s)\}.
\label{eq:gainers}
\end{equation}
Each term contributes one if the constituency strictly gains. A \emph{mutual gain} has \(G(p)=3\). The vote is a separate choice. If \(b_i=1\) denotes an actual yes vote, the institution implements \(p\) when \(\sum_i b_i\geq q\), and otherwise implements \(s\). Majority rule sets \(q=2\) and unanimity sets \(q=3\). The rule counts votes even when a representative votes against its own payoff interest. We record \emph{harmful exclusion} if the implemented plan pays any constituency less than its fallback.

\paragraph{Voting rules.} GPT-4.1 played 18 eight-plan games under both rules. Six games had aligned interests. Each of the 12 conflict menus included exactly one mutual-gain compromise. Within each rule, an agenda setter made an opening offer and two peers made simultaneous counteroffers. Two continuations shared those nominations. One showed the plan nominations alone, the other also showed the accompanying messages. Each produced a new final offer and three binding votes with explicit own-payoff comparisons. The 396 calls produced 72 branches. We averaged the two message conditions within each game and rule before comparing rules. The independent unit was the game.

\paragraph{Offers on fixed histories.} We returned to all 36 full-message histories (18 games under two rules) and elicited four new final offers per history. A two-by-two comparison crossed extra constituency payoff labels with a coalition aid. The aid counted strict gainers, repeated the threshold, and instructed proposers to assume that voters support a gain and oppose a loss. We scored whether the proposer chose its own highest-paying plan among those with enough strict gainers, treating the fallback as feasible. This diagnostic collected 144 offers and no new ballots.

\paragraph{Two experiments with binding votes.} We simplified the menu to the four-plan structure in Table~\ref{tab:menu} and used Qwen3-Next-80B-A3B-Instruct via DeepInfra FP8. The proposer's highest-paying plan benefited two constituencies at the third's expense. Two compromises both improved on the fallback, one paying the proposer more and the other paying the requesting party more. Roles and labels were balanced. A qualification screen required at least 23 of 24 correct decisions and rejection of all six pivotal losing offers. Qwen identified 11 of 12 own-payoff maxima and cast all 12 pivotal ballots correctly. A preceding 30B candidate scored 22/24 and did not advance. The final experiment carried forward the 80B model's qualification.

Each experiment drew 60 fresh games. In each game the party who stood to lose on the two-gainer plan made one public, nonbinding request naming a plan. Two copies of the proposer received the same request, payoff table and unanimity rule. Each made one offer, followed by three fresh binding ballots. Voters saw the request, offer and their own offer-versus-fallback payoffs. They never saw the proposer's display. Generation used categorical tool responses, temperature 0.6 and top-\(p\) 0.95. Each pair used a common proposer seed. Branch order and ballot seeds were counterbalanced.

In the first experiment one proposer received the standard instructions. The other also saw the four counts, their explanation and a repeat of the three-yes threshold. In the final experiment both copies received the same strict-gain definition and the same three-vote reminder. Only the numerical count field differed. We fixed the cases, primary comparison and missing-data rule before collection in each cohort. Every action was observed.

\section{Offers, votes and agreements}

\paragraph{Unanimous agreements with the matched reminder.} In the final 60 games, 26 passed a mutual gain only with counts. One switched the other way (Table~\ref{tab:pairs}). If \(Y_{j,c}\) and \(Y_{j,r}\) indicate enactment of a mutual gain in game \(j\) with counts or with the matched reminder alone, the paired estimate is
\begin{equation}
\widehat{\Delta}=\frac{1}{60}\sum_{j=1}^{60}(Y_{j,c}-Y_{j,r})
=\frac{26-1}{60}=\frac{5}{12}\approx 0.417 .
\label{eq:paired}
\end{equation}
The difference is 41.7 percentage points (conservative 95\% interval 18.8 to 58.5, exact two-sided paired \(p=4.17\times10^{-7}\)). The test assumes that directions of independent discordant pairs are exchangeable under the null. The interval combines simultaneous binomial bounds on the two discordance probabilities. These estimates concern games from this design and model route.

\begin{table}[htbp]\centering\small
\caption{Paired outcomes under unanimity. Each column contains 60 games with fresh votes on both offers. In the right column the threshold reminder and definition were identical across branches.}
\label{tab:pairs}
\begin{tabular}{lrr}\toprule
Outcome within game & Original display comparison & Matched reminder\\\midrule
Passes only with counts & 36 & 26\\
Passes only without counts & 0 & 1\\
Passes in both & 1 & 5\\
Passes in neither & 23 & 28\\\midrule
Total passing without counts & 1/60 & 6/60\\
Total passing with counts & 37/60 & 31/60\\\bottomrule
\end{tabular}
\end{table}

Across the 120 final branches, all 37 mutual-gain offers passed. The requesting party vetoed all 83 offers that would have harmed it. In a descriptive check, we found that this party had asked for its own preferred mutual gain in \emph{every one} of the 60 games. With counts the proposer chose that request 17 times and the other, proposer-favored compromise 14 times. Without counts it chose those plans four and two times, respectively. Both compromises displayed the same count of three.

\begin{figure}[htbp]\centering
\includegraphics[width=\linewidth]{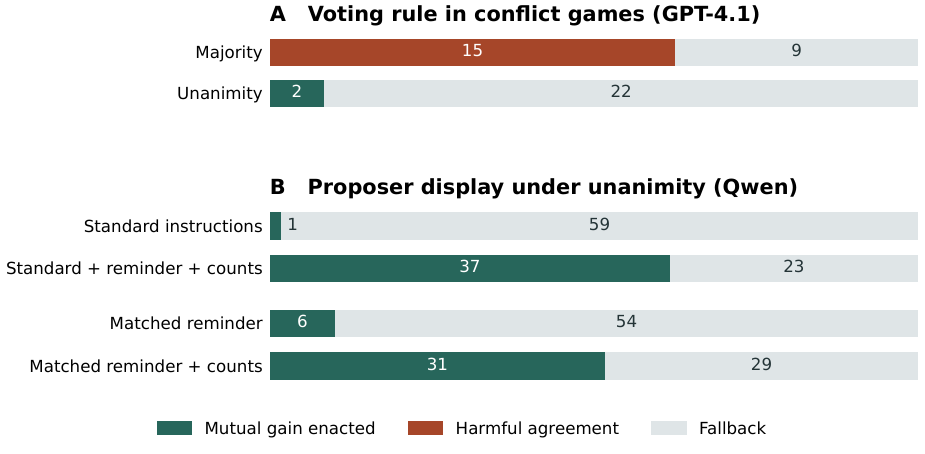}
\caption{Enacted outcomes. Panel A shows 12 GPT-4.1 conflict games with two message conditions per rule. Panel B shows 60 paired Qwen games in each of two separate cohorts. Numbers are branch counts.}\label{fig:outcomes}
\end{figure}

\paragraph{Majority and unanimity.} Majority rule enacted a change in 15 of 24 conflict branches. Each deal cost one constituency points. Unanimity enacted two changes in 24 branches, both mutual gains. All 24 aligned-control branches enacted the dominant plan. Across the full 18 games, unanimity reduced harmful exclusion by 41.7 percentage points (marginal 95\% interval for unanimity minus majority \(-63.9\) to \(-22.2\), adjusted \(p=.0156\)). It raised the minimum realized constituency payoff by 5.11 points on the 0--100 scale (interval 2.72 to 7.61, adjusted \(p=.0078\)). Seeing the peers' messages changed proposer concessions by an estimated \(-4.36\) points (interval \(-12.42\) to \(1.25\)). The marginal intervals use the 2.5th and 97.5th percentiles of the full game-level bootstrap distribution. The exploratory sign-flip tests use a Bonferroni adjustment across four prespecified comparisons.

Applying a three-vote threshold to the existing majority ballots would have blocked all 15 harmful conflict deals, holding the offers and votes fixed. In the actual unanimous conflict games the peers nominated the compromise in zero of 24 counteroffers. Agenda setters put it to a final vote in two of 24 branches. These are descriptive results for the 12 conflict games.

On the fixed histories, the coalition aid increased selection of the proposer's highest-paying feasible plan by 31.9 percentage points, averaging over labels and rules within each game (marginal game-bootstrap interval 18.1 to 45.8, three-test-adjusted exploratory \(p=.00293\)). In a retrospective comparison of the 12 unanimous conflict histories, standard instructions elicited one compromise proposal and the coalition aid elicited nine. These replayed offers received no ballots.

\paragraph{Earlier display comparison.} In the first 60 Qwen games, adding the count display and threshold reminder raised enacted mutual gains from 1 to 37 (+60.0 percentage points, conservative 95\% interval 37.7 to 74.0). All 38 mutual-gain offers passed. The losing party vetoed all 82 offers that would have reduced its payoff.

\section{How much did the counts help?}

Choosing uniformly from the four plans would \emph{propose} a mutual gain half the time in expectation. The matched counts condition proposed one in 31 of 60 games (51.7\%). Always putting the requesting party's plan to a vote would have nominated a mutual gain in all 60 observed cases. Neither benchmark was run as a voting condition. The experiments establish that an arithmetic summary of public payoffs can change proposal selection. A proposer could also succeed by choosing any option marked three. Its choice does not reveal whether it calculated expected votes, attended to the cue, or followed that simple rule.

These games have known payoffs, a finite menu and one final offer per branch. Each study uses the same model in all three roles. Voters receive explicit comparisons of their own payoffs, and the Qwen peer request can only name a plan. Actual ratification of the counts intervention was tested on one model under unanimity. Its effect under majority voting, private information, mixed-model committees or open-ended negotiations remains unmeasured. The two Qwen cohorts used different games and ran at different times. Subtracting their effects would not identify how much the reminder contributed.

The research sequence was adaptive. The earlier missed proposals motivated the fixed-history diagnostic. Its own controls passed, but its combined aid yielded only 30 of 36 highest-paying feasible choices, below the 35 required to advance to a fresh eight-plan study. We then tested the narrower four-plan design. There is no adjustment for comparisons across the wider research program.

The representative in Table~\ref{tab:menu} made a request all three parties could gain from, then voted down an offer that would have cost it nine points. Its vote kept the group's 42 points. Only the proposer could have put the 58-point deal to a vote.

\section*{Data and analysis}
The \href{https://github.com/tofuadmiral/institutional-representation-abm/tree/main/paper3/manuscript}{companion repository} includes the prompts, tool definitions and returned actions for the four studies, game-level tables, and code for the paired estimates and Figure~\ref{fig:outcomes}. Both candidate-model screens are tabulated. The fixed-history study reuses the 18 games from the rule comparison. The four studies' 1,644 calls count executions. Inference uses paired games.

\end{document}